\documentclass[reprint,amsmath,amssymb,showpacs,aps,prr]{revtex4-1}
\usepackage{amsmath,amssymb,amsfonts,bm}
\usepackage{graphicx}
\usepackage{epstopdf}
\usepackage{dcolumn}
\usepackage{mathrsfs}
\usepackage{color}
\usepackage{multirow}
\usepackage{threeparttable}
\usepackage{appendix}
\usepackage[colorlinks=true,linkcolor=blue,citecolor=blue, urlcolor=blue,bookmarks=false]{hyperref}

\begin{document}

\title{State Classification via a Random-Walk-Based Quantum Neural Network}
\author{Lu-Ji Wang}
\affiliation{Institute for Brain Sciences and Kuang Yaming Honors School, Nanjing University, Nanjing 210023, China}
\affiliation{School of Physics, Nanjing University, Nanjing 210093, China}
\author{Jia-Yi Lin}
\affiliation{Institute for Brain Sciences and Kuang Yaming Honors School, Nanjing University, Nanjing 210023, China}
\affiliation{School of Physics, Nanjing University, Nanjing 210093, China}
\author{Shengjun Wu}
\email{sjwu@nju.edu.cn}
\affiliation{Institute for Brain Sciences and Kuang Yaming Honors School, Nanjing University, Nanjing 210023, China}
\affiliation{School of Physics, Nanjing University, Nanjing 210093, China}



\begin{abstract}
    In quantum information technology,
        crucial information is regularly encoded in different quantum states.
    To extract information,
        the identification of one state from the others is inevitable.
    However, if the states are non-orthogonal and unknown,
        this task will become awesomely tricky,
        especially when our resources are also limited.
    Here, we introduce the quantum stochastic neural network (QSNN),
        and show its capability to accomplish the binary discrimination of quantum states.
    After a handful of optimizing iterations,
        the QSNN achieves a success probability close to the theoretical optimum,
        no matter whether the states are pure or mixed.
    Other than binary discrimination,
        the QSNN is also applied to classify an unknown set of states into two types: entangled ones and separable ones.
    After training with four samples, it can classify a number of states with acceptable accuracy.
    Our results suggest that the QSNN has the great potential to process unknown quantum states in quantum information.
\end{abstract}
\maketitle

     Quantum state discrimination identifies a quantum state among an already known set of candidate states.
     It is a key step in many quantum technologies as quantum states are the carriers of information in quantum computing protocols and quantum information processing.
     Although quantum mechanics fundamentally forbids deterministic discrimination of non-orthogonal states, probabilistic methods such as the minimum error discrimination \cite{helstrom1969quantum} and the unambiguous discrimination \cite{ivanovic1987differentiate, dieks1988overlap, peres1988differentiate}, have been developed, and their theoretical best performances have also been found.
     Beyond state discrimination, a special case of classification, one would expect to classify quantum states by more properties in one's own particular interests.
     A well studied example is the classification of states according to whether the state is entangled or not.
     Lots of strategies have been proposed for detecting entanglement, such as using the positive partial transpose (PPT) criterion \cite{horodecki1995violating}, entanglement witnesses \cite{horodecki1996separability, terhal2000bell, lewenstein2000optimization}, and the Clauser-Horne-Shimony-Holt (CHSH) inequality \cite{bell1964einstein,clauser1969proposed}.

    Most of the strategies mentioned above require complete knowledge of the quantum states before carrying out their best practices.
    However, exactly determining all the states is in principle forbidden, unless infinite copies of the states are provided.
    In Ref. \cite{massar2005optimal}, Massar and Popescu have addressed this topic, known as the state estimation, and proved that the optimal mean fidelity for two-level system state estimation is $\frac{N+1}{N+2}$, where $N$ is the number of identically prepared copies of the state.
    The optimal mean fidelity is less than $1$ and tends towards $1$ as the number of copies $N$ tends to infinity.
    Then, in Ref. \cite{bruss1999optimal} the authors extended the result to finite dimensional quantum systems in pure states.
    All these results indicate that the pre-processing of quantum state estimation before carrying out the state discrimination or classification will introduce extra errors.
    In addition, state estimation often involves a quantum information technology called state tomography, which is sometimes prohibitively expensive to perform on an unknown state \cite{paris2004quantum}.
    In the states classification task, it also becomes practically impossible to do a tomography for each state, because the number of states waiting to be classified can be extremely large.
    In addition to the extra errors introduced by the pre-processing and the expensive tomography, there is another difficulty for these traditional strategies.
    That is, even if we exactly know the quantum states to be identified or classified, the optimal measurement is hard to be analytically developed in the fashion of traditional strategies, when more than two states are involved \cite{bergou2010discrimination, barnett2009quantum}.

    Facing the above difficulties, we hope there will be a new strategy for state discrimination and classification.
    Recently, there has been a rising trend of using machine learning methods to fully exploit the inherent advantages of quantum technologies \cite{sarma2019machine, carleo2019machine}.
    Among these studies, the classification problem has received a great deal of attention, and some quantum classifiers have shown excellent performance \cite{schuld2020circuit, li2022recent}.
    In addition, as a fusion of deep learning and quantum computation, quantum neural networks \cite{biamonte2017quantum, dunjko2018machine} have been proved to be effective and almost irreplaceable in many quantum tasks, such as quantum operation implementations \cite{he2021variational,beer2020training,steinbrecher2019quantum}, many-body system simulations \cite{carleo2017solving, gao2017efficient}, and quantum autoencoders \cite{wan2017quantum,steinbrecher2019quantum, bondarenko2020quantum}.

    There are also several successful attempts of classifying quantum states with quantum neural networks. Chen \emph{et al.} \cite{chen2021universal} utilized a hybrid approach to learn the design of a quantum circuit to distinguish between two families of non-orthogonal quantum states and generalizes its ability to previously unseen quantum data.
    This approach was further extended to noisy devices by Patterson \emph{et al.} \cite{patterson2021quantum}
    Cong \emph{et al.} \cite{cong2019quantum} introduced a quantum convolutional neural network accurately recognizing quantum phases.
    Those neural networks can all be implemented on near-term devices.
    Dalla Pozza \emph{et al.} \cite{dalla2020quantum} proposed a quantum network with state discrimination capability based on an open quantum system.
    A tightly related protocol was then experimentally implemented by Laneve \emph{et al.} \cite{laneve2021experimental}, which provided a novel approach to multi-state discrimination.

    Inspired by these works, in the Letter we introduce a new kind of quantum neural network, i.e., the quantum stochastic neural network (QSNN), to complete quantum state discrimination and classification tasks. Our network is based on quantum stochastic walks (QSWs) \cite{whitfield2010quantum}, which have been theoretically proposed and experimentally implemented to simulate the associative memory of Hopfield neural networks \cite{schuld2014quantum, tang2019experimental}. Therefore, it is worthwhile to explore the power of quantum walks in building general quantum neural networks.

   \emph{Approach}.
    A classical random walk describes the probabilistic motion of a walker over a graph.
    Farhi and Gutmann \cite{farhi1998quantum} generalized classical random walks into quantum versions, i.e., continuous-time quantum walks (CTQWs).
    QSWs are further generalizations of CTQWs by introducing the decoherence, so that both classical and quantum random walks can be described by them.
    The state of the walker is described by a density matrix $\rho$, and evolves as \cite{kossakowski1972quantum, lindblad1976generators, gorini1976completely}
    \begin{equation}\label{eq:master equation}
        \frac{d\rho}{dt}=-i[H,\rho]+\sum_{k} \left(L_k\rho L_k^{\dagger}-\frac{1}{2}\{L^\dagger_k L_k,\rho\} \right),
    \end{equation}
    where $H$ is the Hamiltonian and $L_k$ is a Lindblad operator.

\begin{figure}
  \centering
  \includegraphics[width=5cm]{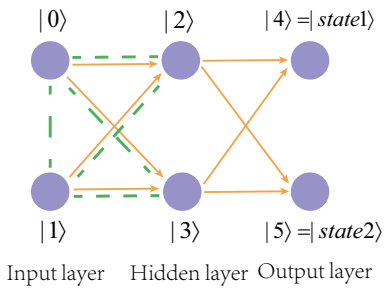}\\
  \caption{It is the graph representation of the quantum stochastic neural network (QSNN) used for quantum state binary discrimination.
  The state of the QSNN can be represented by a density matrix in the $N=6$ dimensional Hilbert space constituted by an orthogonal basis $\{|i\rangle\}_{i=0}^5$.
  The state of the two neurons of the input layer is initialized as the state of a 2-level input quantum state.
  Two neurons in the output layer correspond to the 2 labels $state1$ and $state2$ that distinguish two different states.
  The vertices are decoherently connected (orange lines with arrows) by Lindblad operators and coherently connected (green dashed lines) by Hamiltonian elements. }\label{fig:2state}
\end{figure}

    Our QSNN is based on QSWs (implied by Eq.~(\ref{eq:master equation})) rather than CTQWs, because we need to introduce the decoherence to the QSNN to simulate the forward propagation of probabilities in classical networks.
    The state of the QSNN is described by a density matrix $\rho=\sum_{ij}\rho_{ij}|i\rangle\langle j|$ in the $N$ dimensional Hilbert space,
        where $\{|i\rangle\}_{i=0}^{N-1}$ is an orthogonal basis,
        and each basis state $|i\rangle$ corresponds to a vertex of the graph.
    The vertices can be seen as neurons of the QSNN.
    As an example, a QSNN consisting of three layers, $6$ neurons is shown in Fig. \ref{fig:2state}.
    The state of the QSNN evolves according to Eq. (\ref{eq:master equation}), where the Hamiltonian $H$ and the Lindblad operators $L_k$ respectively determine the coherent and decoherent part of the dynamic.
    In our approach, we use the Hamiltonian $H=\sum_{ij}h_{ij}|i\rangle\langle j|$ to characterize the coherent transmission between neurons.
    The coefficients $h_{ij}$ are complex numbers with the requirement $h_{ij}=h_{ji}^*$ to ensure the hermiticity of the Hamiltonian in general.
    However, real coefficients $h_{ij}\in\mathbb{R}$ are sufficient for the tasks of our interests and we don't consider the coupling of a neuron to itself.
    Thus, the Hamiltonian is written as
    \begin{equation}\label{eq:hamiltonian}
    \begin{aligned}
        H=\sum_{ij}h_{ij}|i\rangle\langle j| = \sum_{i<j}h_{ij}(|i\rangle\langle j|+|j\rangle\langle i|).
    \end{aligned}
    \end{equation}
    The Lindblad operator used by us in Eq. (\ref{eq:master equation}) is
    \begin{equation}
    \begin{aligned}\label{eq:lindblad}
        L_k\rightarrow L_{ij}=\gamma_{ij}|i\rangle\langle j |,
    \end{aligned}
    \end{equation}
    which simulates the decoherent (typically one-way) transmission from the $j$th neuron to the $i$th neuron.
    The coefficient $\gamma_{ij}$ that characterizes the dissipation rate is a real number in general.
    We group the coefficients in the Hamiltonian as a single vector $\bm{h}=(h_1, h_2, \cdots, h_k, \cdots)$ and the coefficients in the Lindblad operators as an another vector $\bm{\gamma}=(\gamma_1, \gamma_2, \cdots, \gamma_k, \cdots)$.
    The coefficient vectors $\bm{h}$ and $\bm{\gamma}$ are the parameters of the QSNN that need to be optimized.
    In this study, the dissipation and the Hamiltonian couplings only exist between certain neurons, which is discussed in more detail in the Supplementary Material (SM) I.A.
    As shown in Fig.~\ref{fig:2state}, the Lindblad operators (the orange lines with arrows) only connect the neurons in the adjacent layers to transfer the probability amplitude from one layer to the other, so that probability converges in the output layer.
    And the Hamiltonian couplings (the green dashed lines) only exist between some neurons in the input and hidden layer to perform a non-directional transmission.

    The QSNN is initialized by encoding the state to be classified in the state of the input layer neurons of the network.
    To be specific, if we use an $N$-dimensional QSNN with $n$ input layer neurons to classify $n$-level quantum states $\rho$, the state of the network would be initialized as $\rho_{\text{in}}=\rho\oplus 0_{N-n, N-n}$, where $0_{i, j}$ represents an $i\times j$ zero matrix.
    Then, the network evolves according to Eq. (\ref{eq:master equation}) for a duration $T$ from its initial state $\rho_\text{in}$ and gives the final state $\rho_{\text{out}}^s$ for the $s$th input.
    The evolution time $T$ is considered dimensionless (it is of dimension $1/\gamma$ actually, where $\gamma$ is a typical value of the coupling parameters $h_k$ in Hamiltonian or the dissipation rates $\gamma_k$ in Lindblad operators).
    The final state describes the probability that the walker is on each vertex (neuron) at time $T$.
    The probability converges in the output layer due to the existence of the one-way decoherent transmission.
    We correspond output neurons to the labels that distinguish all kinds of different states one by one.
    For the example of the QSNN shown in Fig. \ref{fig:2state}, if $\rho_\text{out}^s=|N-2\rangle\langle N-2|$ ($\rho_\text{out}^s=|N-1\rangle\langle N-1|$), we say the unknown quantum state belongs to class 1 (class 2).
    Hidden layers should be set according to tasks, and a single hidden layer is sufficient in the tasks of our interests [details in SM I.B].

    In the training process, we firstly draw an already labeled sample state $\rho_{\text{in}}^s$ together with its label $l^s\in\{N-2, N-1\}$ from a training set $\{(\rho_{\text{in}}^s, l^s)\}_{s=1}^M$, where $M$ is the number of samples in the training set.
    Then, performing a projective measurement $\varOmega^s = |l^s\rangle\langle l^s|$ on the final state $\rho_\text{out}^s$ of the network gives the success probability
    \begin{equation}\label{eq:success probability}
    P_{\text{N}}^s = \text{Tr}(\rho_{\text{out}}^s\varOmega^s)
    \end{equation}
    that the QSNN gives the desired output corresponding to the $s$th sample.
    We can design the specific forms of the loss function
    \begin{equation}
    \mbox{Loss}=\mbox{Loss}\left(\bm{h}, \bm{\gamma},\{(\rho^s_{\text{in}}, l^s)\}\right) = \sum_s w_s f(\rho_{\text{out}}^s, \varOmega^s)
    \end{equation}
    according to different tasks, where $w_s$ is a weight on the sample $s$, and $f$ is the sample-wise loss.
    The loss function should be designed such that minimizing it (by gradient descent introduced in SM II) leads to the result that the QSNN classifies states correctly.

\begin{figure}[ht]
  \centering
  \includegraphics[width=8cm]{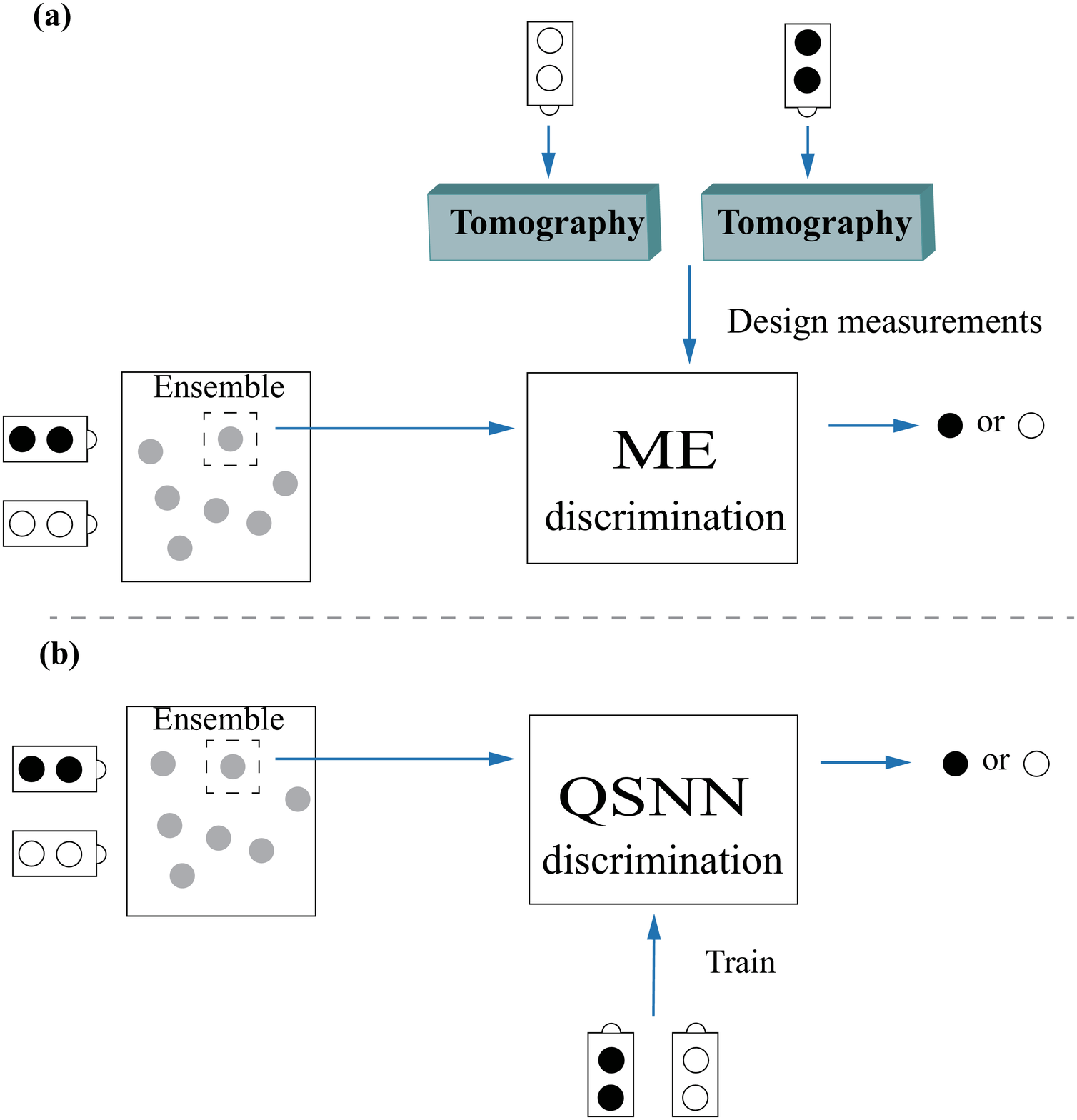}
  \caption{The flow charts of quantum state binary discrimination task.
  Black and white dots respectively represent the unknown quantum states from two different devices, and they become indistinguishable (gray dots) when mixed in an ensemble.
  The task is that we randomly pick one state from the ensemble and determine which device the state is coming from.
  (a) The minimum error discrimination. Before the discrimination, a pre-processing of quantum state tomography is often needed to get the complete information of the unknown states.
  (b) The QSNN discrimination. There is no need to obtain any prior information of quantum states through tomography in our approach. The network can discriminate quantum states after being trained with some labeled samples.
  }\label{fig:discriminationflow}
\end{figure}

    \emph{Results -- Quantum State Binary Discrimination.}
    The general processes of the minimum error (ME) discrimination and our QSNN discrimination are respectively shown in Fig. \ref{fig:discriminationflow}.
    There is an ensemble where quantum states are respectively prepared by two devices.
    The two kinds of quantum states in the ensemble are unknown, i.e., not well-defined mathematically.
    We randomly pick one of the states from the ensemble.
    The task, called quantum state binary discrimination, is to determine which kind of state we have picked.
    The quantum states $\rho_1$ and $\rho_2$ are prepared with prior probabilities $w_1$ and $w_2$ ($w_1 + w_2 = 1$), respectively.

    As shown in  Fig. \ref{fig:discriminationflow}(a), before discriminating two unknown states in the ensemble with ME discrimination, a pre-processing of quantum state tomography is often needed, so that the appropriate measurement can be set up.
    It is not trivial to find the ME measurement set-up in general cases, but for the quantum state binary discrimination,
    the minimum error probability is given analytically by Helstrom \cite{helstrom1969quantum} and Holevo \cite{holevo2011probabilistic}, which is called the Helstrom bound:
    \begin{equation}\label{helstrom bound}
    P_\text{H}^{\text{error}} = \frac{1}{2}(1-\text{Tr}|w_2\rho_2-w_1\rho_1|).
    \end{equation}
    Then, the success probability of the ME discrimination is given as
    \begin{equation}
    P_{\text{H}} = 1-P_H^{error}.
    \end{equation}

    However, as shown in Fig. \ref{fig:discriminationflow}(b), tomography is not required before our QSNN discrimination. The QSNN can discriminate quantum states by learning from some labeled samples, without knowing the mathematical expressions of these states.
    In order to complete the quantum state binary discrimination task, the QSNN is constructed by 6 neurons divided into three layers, as shown in Fig. \ref{fig:2state}.
    Some discussions about the topology of the network are given in SM I.
    There are only two sample states in the training set, namely $\rho_1$ and $\rho_2$.
    They are labeled $state\ 1$ and $state\ 2$, and fed into the QSNN with the prior probability $w_1$ and $w_2$, respectively.
    Thus, according to Eq. (\ref{eq:success probability}), the average success probability that the QSNN correctly gives the labels of the two input states is given as
    \begin{equation}
        P_{\text{N}}=\sum_{s=1}^{2}w_s\text{Tr}(\rho_{\text{out}}^s\varOmega^s).
    \end{equation}
    Then, the loss function is defined as the distance between 1 and the average success probability, that is,
    \begin{equation}\label{eq:discrimination loss}
    \mbox{Loss}=1-P_{\text{N}}.
    \end{equation}
    In our simulation, we choose $w_1 = w_2 = 0.5$.
    To show the performance of our approach, we compare the success probability of the QSNN ($P_{\text{N}}$) and that of the ME discrimination ($P_{\text{H}}$).

\begin{figure}[h]
  \centering
  \includegraphics[width=9.0cm]{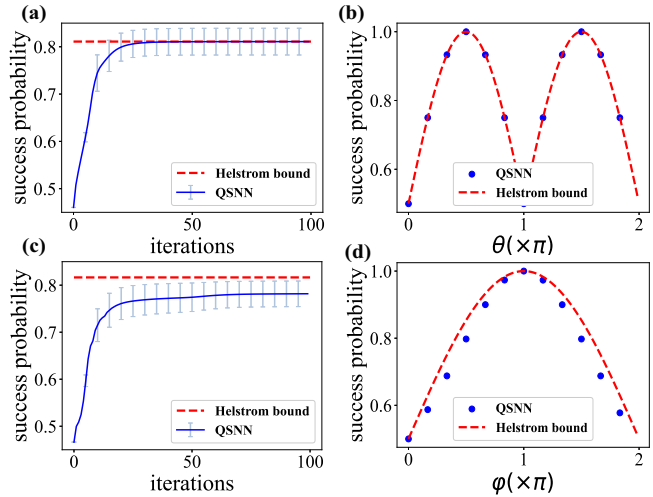}\\
  \caption{(a) The blue curve represents the average success probability of the QSNN for the discrimination of different quantum state pairs $(|\psi_0\rangle, |\psi_{\theta}\rangle)$ with real amplitudes.
  The red dashed line shows the average optimal theoretical value given by the minimum error discrimination.
  The results are given by taking the average over the success probabilities corresponding to all state pairs with different values of $\theta$.
  The error bars are plotted from the variances.
  The average success probability rises rapidly, and after about 30 iterations, it achieves the Helstrom bound.
  (b) The optimal success probabilities of the QSNN in discriminating $|\psi_0\rangle$ and $|\psi_\theta\rangle$ with each $\theta$ are drawn as blue dots. The Helstrom bound is shown as the red dashed line.
  Here, (c) and (d) include the results similar to (a) and (c), respectively. They are given by replacing the state represented by Eq. (\ref{eq:real state}) with that represented by Eq. (\ref{eq:complex state}).
  The QSNN can achieve a success probability of no less than 91\% of the theoretical optimum.}\label{fig:pure}

 \end{figure}

    First, without loss of generality, we consider the quantum states
    \begin{equation}\label{eq:real state}
    |\psi_{\theta}\rangle=\cos{\theta}|0\rangle+\sin{\theta}|1\rangle
    \end{equation}
    in the real vector space. Here $\{|0\rangle, |1\rangle\}$ constitutes an orthogonal basis.
    The specific expressions of the quantum states selected here are intended only to mathematically evaluate the performance of our model.
    The states we want to discriminate against are $|\psi_0\rangle$ and $|\psi_{\theta}\rangle$, so our training set is $\{(|\psi_0\rangle, state1), (|\psi_{\theta}\rangle, state2)\}$.
    In our simulation, we train the QSNN with different training sets separately, where $\theta=0, \frac{\pi}{6}, \frac{2\pi}{6}, \frac{3\pi}{6}, \cdots, \frac{11\pi}{6}$.
    The average success probability of the QSNN (blue curve) for the discrimination of 12 quantum state pairs ($|\psi_0\rangle$, $|\psi_{\theta}\rangle$) increases with the number of iterations used in the training procedure, as shown in Fig. \ref{fig:pure}(a).
    Our QSNN approximately achieves the optimal theoretical bound of the success probability named as Helstrom bound (red dashed line) after about 30 iterations.
    The optimal success probability of the QSNN for each training set with different $\theta$ is shown as a blue dot in Fig. \ref{fig:pure}(b).
    Each of them approximates to the Helstrom bound well.

    Second, we consider the quantum states
    \begin{equation}\label{eq:complex state}
    |\psi_{\varphi}\rangle=\frac{\sqrt{2}}{2}(|0\rangle+e^{i\varphi}|1\rangle),
    \end{equation}
    which are in a complex space.
    Similarly, we train the QSNN to discriminate states $|\psi_0\rangle$ and $|\psi_{\varphi}\rangle$ with the different training sets $\{(|\psi_0\rangle, state1), (|\psi_{\varphi}\rangle, state2)\}$, which are different in $\varphi=0, \frac{\pi}{6}, \frac{2\pi}{6}, \frac{3\pi}{6}, \cdots, \frac{11\pi}{6}$.
    The result shown in Fig. \ref{fig:pure}(c) is also the average success probabilities for all training sets.
    We can see a certain gap between the success probability of the QSNN and the Helstrom bound.
    Fig. \ref{fig:pure}(d) indicates that the trained QSNN does not perform well in discriminating the quantum states with complex amplitudes.
    Even so, the discrimination result given by the optimized QSNN is still referential, because it can achieve a success probability of no less than 91\% of the theoretical optimum.

    In summary, the QSNN can complete binary discrimination of unknown quantum states without tomography.
    It can be trained to its optimum after a few iterations and used to discriminate the states with a single detection.
    The optimized QSNN can achieve a success probability close to the Helstrom bound for both pure and mixed state (displayed in SM III) discrimination.
    Our model also works when the dimension of quantum states to be discriminated against is greater than 2, and we show the simulation result of the 3-qubit case in Fig. S8 of the SM.
    In addition, our approach is not limited by the number of states to be discriminated against theoretically, as shown in SM~IV.
    In opposite, in general, if the number of the states is greater than two, the optimal measurement will be hard to be analytically developed in the fashion of traditional strategies.

\emph{Results -- Classification of Entangled and Separable States.}
    Entanglement is a primary feature of quantum mechanics, which is also considered with a resource.
    As the carrier of entanglement, entangled quantum states are costly to produce.
    Therefore, determining whether the given state is entangled or not is an important topic in quantum information theory.
    Now, some unknown separable states and entangled quantum states are mixed in an ensemble.
    We only know that they are prepared by two devices with an equal prior probability, without knowing their mathematical expressions.
    The task we consider in the following is to determine which category each quantum state belongs to. This is a classification task.

    Several traditional strategies have been proposed to complete this task.
    For example, the positive partial transpose (PPT) criterion \cite{horodecki1995violating} is both sufficient and necessary for 2-qubit entanglement detection with the requirement of quantum tomography (see Fig. S9 of the SM).
    The CHSH inequality \cite{bell1964einstein, clauser1969proposed} is also an attractive strategy because it only needs partial information of the quantum states (see Fig. S9 of SM). However, on the one hand, multiple measurements are still required.
    On the other hand, using fixed measurements, the CHSH inequality can't detect all entangled states in a set of states.
    To optimize the accuracy of the entanglement detection using the CHSH inequality, classical artificial neural networks combined with machine learning techniques have been proposed in Refs. \cite{gao2018experimental,ma2018transforming}.
    They construct a quantum state classifier and it achieves the accuracy of the classification to near unity, but multiple measurements are still required.
    To avoid multiple measurements or state tomography, we train a QSNN to be a quantum classifier.

    In order to evaluate our approach more clearly, we select an unknown set of Werner-like states in our simulation.
    Each state is of the form
    \begin{equation}\label{eq:werner like states}
     \rho = p|\varPsi\rangle\langle\Psi|+(1-p)\frac{I}{4},
    \end{equation}
    where $|\varPsi\rangle=(U_1\otimes U_2)|\psi_+\rangle$ and the real coefficient $p\in[0, 1]$.
    $U_1$ and $U_2$ are two unknown local unitaries acting on the Bell states $|\psi_+\rangle=\frac{\sqrt{2}}{2}(|01\rangle+|10\rangle)$.
    The state $\rho$ can be regarded as the convex combination of an unknown maximally-entangled state and the maximally mixed state $\frac{I}{4}$.
    The quantum state $\rho$ is separable when $p\leq\frac{1}{3}$ and it is entangled otherwise.

    In our simulation, there are four neurons in the input layer and four in the hidden layer.
    Two neurons in the output layer correspond to two labels $l^s$: reparable $S$ and entangled $E$.
    To be more specific, if the final state $\rho_{\text{out}}$ is $|S\rangle\langle S|$ ($|E\rangle\langle E|$), the trained QSNN shows that the input quantum state is separable (entangled).
    Performing a corresponding measurement $\varOmega^s=|l^s\rangle\langle l^s|$ on the final state $\rho_{\text{out}}^s$ gives the probability that the QSNN gives the label of the $s$th input state correctly.
    The loss function is defined as the mean error probability of all the $M$ training samples
    \begin{equation}
    \mbox{Loss}=1-\frac{1}{M}\sum_{s=1}^M\text{Tr}(\rho_{\text{out}}^s\varOmega^s).
    \end{equation}

\begin{figure}[h]
  \centering
  \includegraphics[width=9cm]{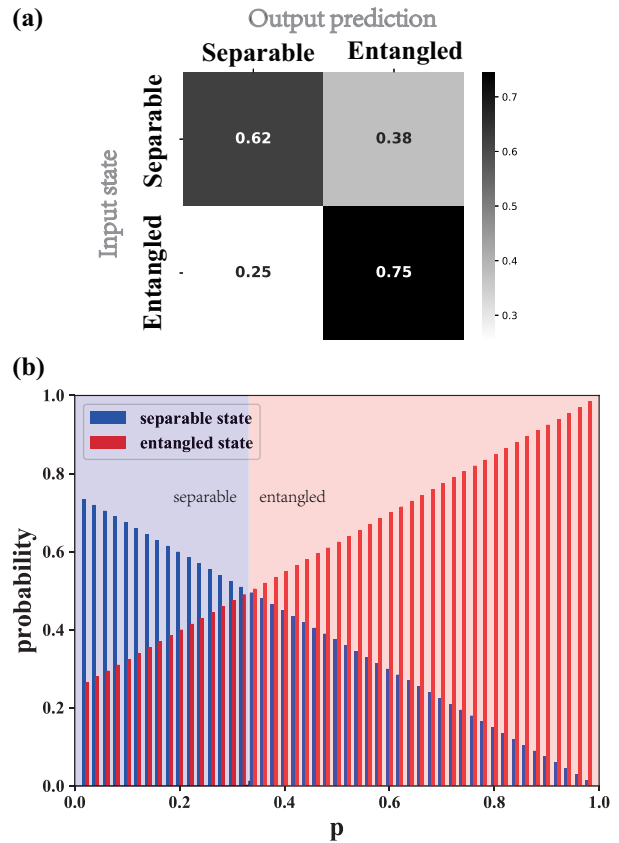}
  \caption{(a) The classification confusion matrix of the QSNN.
  The numbers in it are the mean success and error probabilities of the QSNN in classifying the 49 samples represented by Eq.~(\ref{eq:werner like states}) with $p\in\{0.02\cdot n\}_{n=1}^{49}$.
  The QSNN performs better when the input states are entangled.
  (b) The classification result of each state with a specific value of $p$. The bar chart shows the probabilities that the trained network identifies the 49 input states as entangled states (red bars) and separable states (blue bars).
  The whole chart is divided by $p=\frac{1}{3}$ into the light blue area (the state is separate) and the light red area (the state is entangled).
  The success probabilities are always higher than the error probabilities.}\label{fig:separate}
\end{figure}

    For the example of $U_1=\sigma_z$, $U_2=I$, we numerically give the classification results.
    We only use $M=4$ training samples, where $p\in\{0, 0.2, 0.4, 0.8\}$, while use 49 states with $p\in\{0.02\cdot n\}_{n=1}^{49}$ to evaluate the performance of the trained QSNN in the simulation.
    The classification confusion matrix for these 49 states is shown in Fig.~\ref{fig:separate}(a).
    The trained QSNN identifies separable states successfully with the probability of 0.62 and identifies entangled states successfully with the probability of 0.75.
    The probability of one kind of the wrong classifications, i.e., the trained network identifies a separable state as an entangled one, is 0.38.
    The probability of the other kind of wrong classifications is 0.25.

    The probabilities that each quantum state with a specific value of $p\in\{0.02\cdot n\}_{n=1}^{49}$ is identified as a separable state (blue bar) and an entangled state (red bar) are shown in Fig. \ref{fig:separate}(b).
    When the input states are entangled at $1/3<p\leq1$ (light red region), the red bars are always longer than blue bars.
    It means that the success probabilities are always higher than the error probabilities, which is also true when $0\leq p \leq 1/3$ (light blue region).

    In summary, when given an unknown quantum state set, the QSNN can be trained by several labeled states and used to classify the others.
    Although the QSNN becomes confused about the states at the boundary $p=1/3$, the success probability is always higher than the error probability for each state.
    What's more, the trained QSNN can give the classification result only using a single detection for each state.
    Some details about the parameters and hyperparameters of the QSNN are given in SM~V.

    In this work, we have introduced the quantum stochastic neural network (QSNN), based on quantum stochastic walks. When combined with machine learning, it can be trained to become a quantum state classifier.

    If one wants to classify an ensemble only containing two unknown quantum states, the classification task becomes binary discrimination.
    The QSNN doesn't need any information of the candidate states in advance, so it avoids the experimentally expensive quantum state tomography used in the traditional minimum error discrimination.
    We have benchmarked the QSNN's performance in the quantum state binary discrimination tasks with numerical simulation.
    The success probability of the QSNN turned out to be very close to the theoretical optimal success probability, i.e., the Helstrom bound.

    When given an unknown ensemble containing states from two different families, the QSNN can be trained to classify them into those two families.
    We show an example of classifying Werner-like states according to whether they are entangled or not.
    For all those states, the trained QSNN is always more likely to classify it into the correct family with a single detection.
    The optimal performance of the QSNN can be achieved with only four training samples, while avoiding the state tomography and multiple measurements required by other classification methods such as using the PPT criteria or the CHSH inequality.
    This also suggests that our approach may reduce the consumption of resources compared to traditional methods.

    All the present results have shown the potential of the QSNN as a general-purpose quantum classifier, which may be helpful in various quantum machine learning models, such as the quantum adversarial generative networks \cite{dallaire2018quantum}.

\emph{Acknowledgments.}
This work is supported by the National Key R\&D Program of China (Grant No. 2017YFA0303703)
and the National Natural Science Foundation of China (Grant No. 12175104).


\end{document}


\title{Supplementary Material for \\
``State Classification via a Random-Walk-Based Quantum Neural Network"}
\author{Lu-Ji Wang}
\affiliation{Institute for Brain Sciences and Kuang Yaming Honors School, Nanjing University, Nanjing 210023, China}
\affiliation{School of Physics, Nanjing University, Nanjing 210093, China}
\author{Jia-Yi Lin}
\affiliation{Institute for Brain Sciences and Kuang Yaming Honors School, Nanjing University, Nanjing 210023, China}
\affiliation{School of Physics, Nanjing University, Nanjing 210093, China}
\author{Shengjun Wu}
\affiliation{Institute for Brain Sciences and Kuang Yaming Honors School, Nanjing University, Nanjing 210023, China}
\affiliation{School of Physics, Nanjing University, Nanjing 210093, China}
\maketitle


\section{Topology of QSNNs}\label{sec:topology}
    \subsection{Absence of the Hamiltonian coupling}\label{subsec:topology link}

    As described by Eq. (2) of the main text, Hamiltonian couplings $h_{ij}$ can exist between any two neurons ($|i\rangle$ and $|j\rangle$) of the QSNN in general.
    However, in the tasks of our interests, the couplings $h_{ij}$ only exist between neurons in the input and hidden layer, as the green dashed lines shown in Fig.~1 of the main text.
    There are only Lindblad operators (orange lines with arrows) between the neurons in the hidden layer and output layer.
    Because we expect the probability transmission between the hidden layer and output layer is one-way, so that probability converges in the output layer.
    But the Hamiltonian performs a non-directional transmission between the connected neurons.
    So, the existence of the couplings $h_{ij}$ between the hidden layer and output layer might slow down the convergence of probability at the output layer, which further makes the QSNN difficult to be trained.
\begin{figure}[h]
  \centering
  \includegraphics[width=9.5cm]{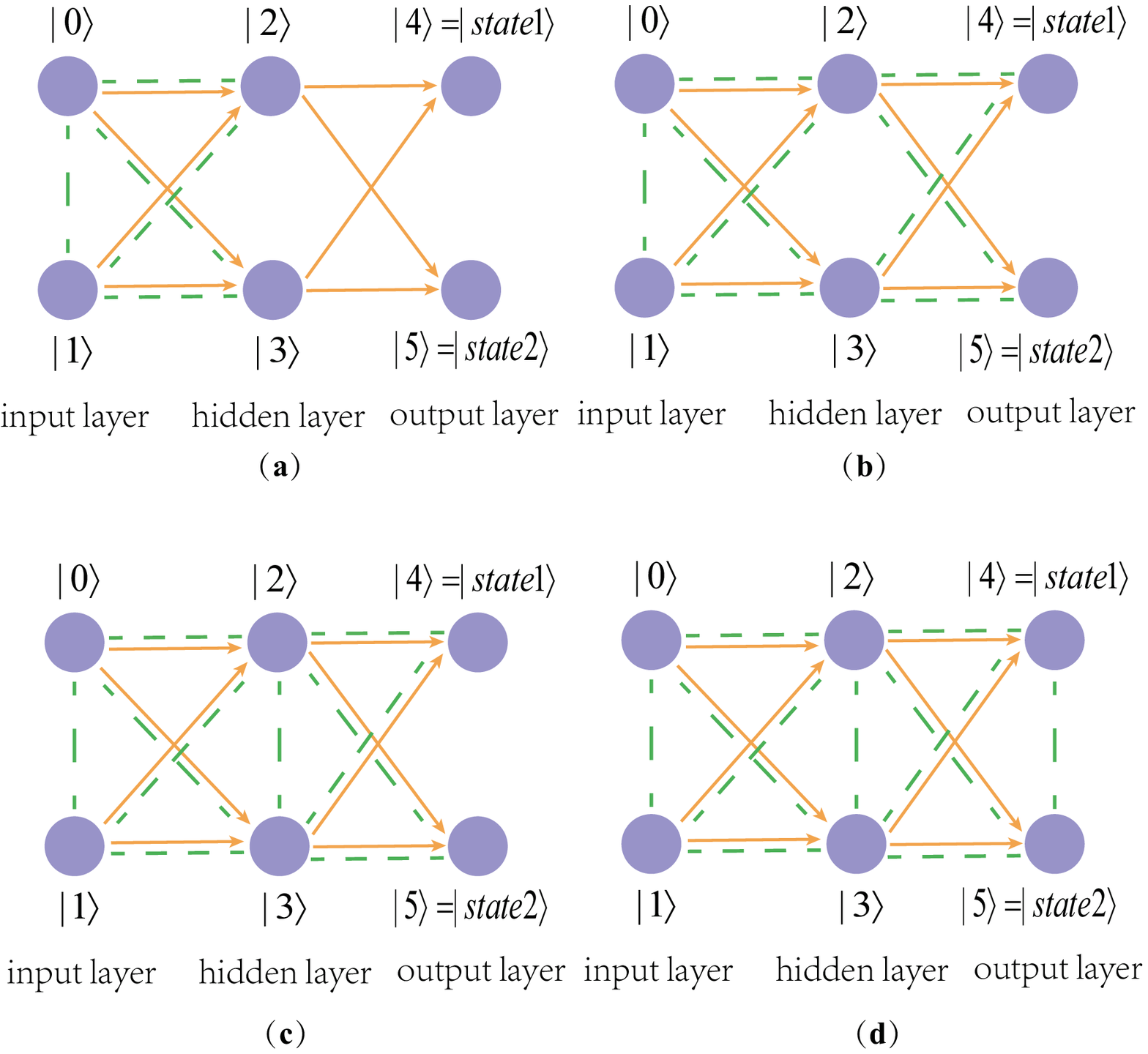}
  \caption{(a), (b), (c) and (d) are the graph representations of the four QSNNs used for quantum state binary discrimination.
  The state of a QSNN can be represented by a density matrix in the $N=6$ dimensional Hilbert space constituted by an orthogonal basis $\{|i\rangle\}_{i=0}^5$.
  The vertices (neurons) are decoherently connected by Lindblad operators (orange lines with arrows) and coherently connected by Hamiltonian elements (green dashed lines).}\label{fig:4model}
\end{figure}

\begin{figure}[h]
  \centering
  \includegraphics[width=8.5cm]{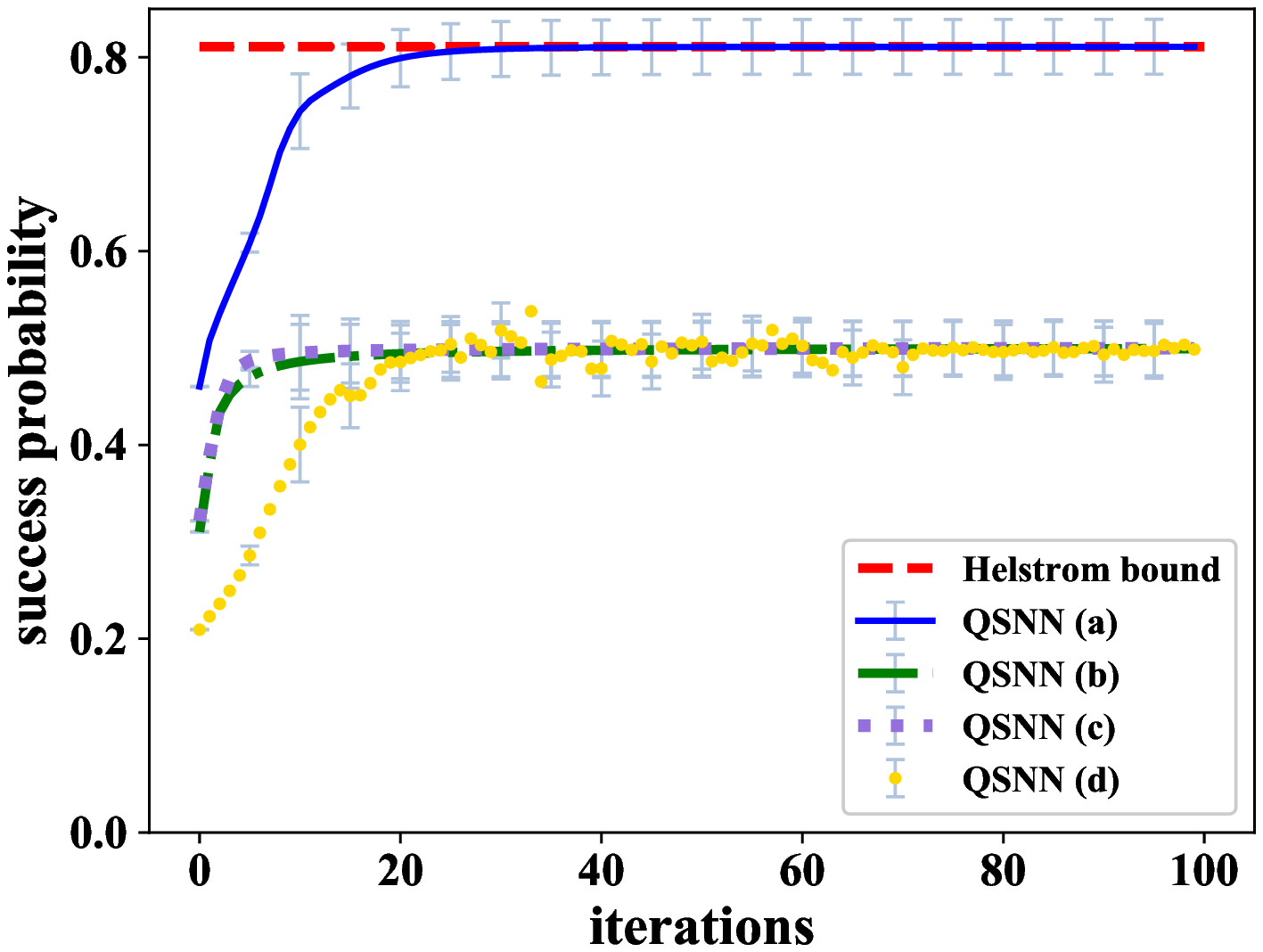}
  \caption{The training performances of the 4 QSNNs (shown in FIG.~\ref{fig:4model}) in the discrimination of quantum state pairs $(|\psi_0\rangle, |\psi_{\theta}\rangle)$.
  The results are given by taking the average over the success probabilities of all training sets with different $\theta$.
  The red dashed line shows the average optimal theoretical value given by the minimum error discrimination.
  The other lines in different colors and shapes represent the average success probability of the 4 QSNNs.
  The error bars are plotted from the variances.
  We can see the QSNN (a), i.e. the QSNN we used in the main text, performs best.
  The success probabilities of other QSNNs are about 0.5, which are terrible performances.}\label{fig:comparison}
\end{figure}

    Then, we give the comparison between the training performances of the four QSNNs (shown in FIG.~\ref{fig:4model}) here.
    We train them to complete the same task as that described in the main text \emph{Results –- Quantum State Binary Discrimination}, i.e., the discrimination between two states represented by
    \begin{equation*}
    |\psi_{\theta}\rangle=\cos{\theta}|0\rangle+\sin{\theta}|1\rangle.
    \end{equation*}
    The training set is $\{(|\psi_0\rangle, state1), (|\psi_{\theta}\rangle, state2)\}$.
    We train the QSNNs with the different training sets separately, where $\theta=0, \frac{\pi}{6}, \frac{2\pi}{6}, \frac{3\pi}{6}, \cdots, \frac{11\pi}{6}$, and average over the results for these training sets.
    FIG. \ref{fig:comparison} shows the training performance of the 4 QSNNs.
    The QSNN shown as FIG. \ref{fig:4model}(a) performs best, and the other QSNNs all perform poorly.

    In addition to the consideration in terms of training performance, we also have to consider the resource consumption of the model.
    The more parameters that need to be optimized, the more resources we consume.

    In summary, the Hamiltonian couplings between the hidden layer and the output layer do not improve the training performance of the network but consume resources, so we use the QSNN shown as FIG.~\ref{fig:4model}(a) without that.

    \subsection{Hidden layer}\label{subsec:number of hidden layer}
    In traditional artificial neural networks, a hidden layer is a layer located between the input and output layer, where neurons take in a set of weighted inputs and produce an output through an activation function.
    The number and size of hidden layers can sometimes affect the performance of the network.
    However, more hidden layers or more hidden layer neurons do not necessarily lead to better model performance.
    One usually selects the appropriate number and size of hidden layers empirically based on the specific task.
    In our approach, we also follow this convention.
    For greater clarity, we use $\mathcal{I}-\mathcal{H}-\mathcal{O}$ to describe the structure of a QSNN, where $\mathcal{I}$, $\mathcal{H}$, and $\mathcal{O}$ are the numbers of neurons in the input layer, hidden layer(s), and output layer, respectively.
    For example, the QSNN we used in the state binary discrimination task is a 2-2-2 network as shown in Fig. 1 of the main text.
    Then, we give some empirical evidence to show that the QSNNs used in the main text are appropriate for our tasks, if both resource consumption and model performance are considered.

  \begin{figure}[ht]
  \centering
  \includegraphics[width=8.5cm]{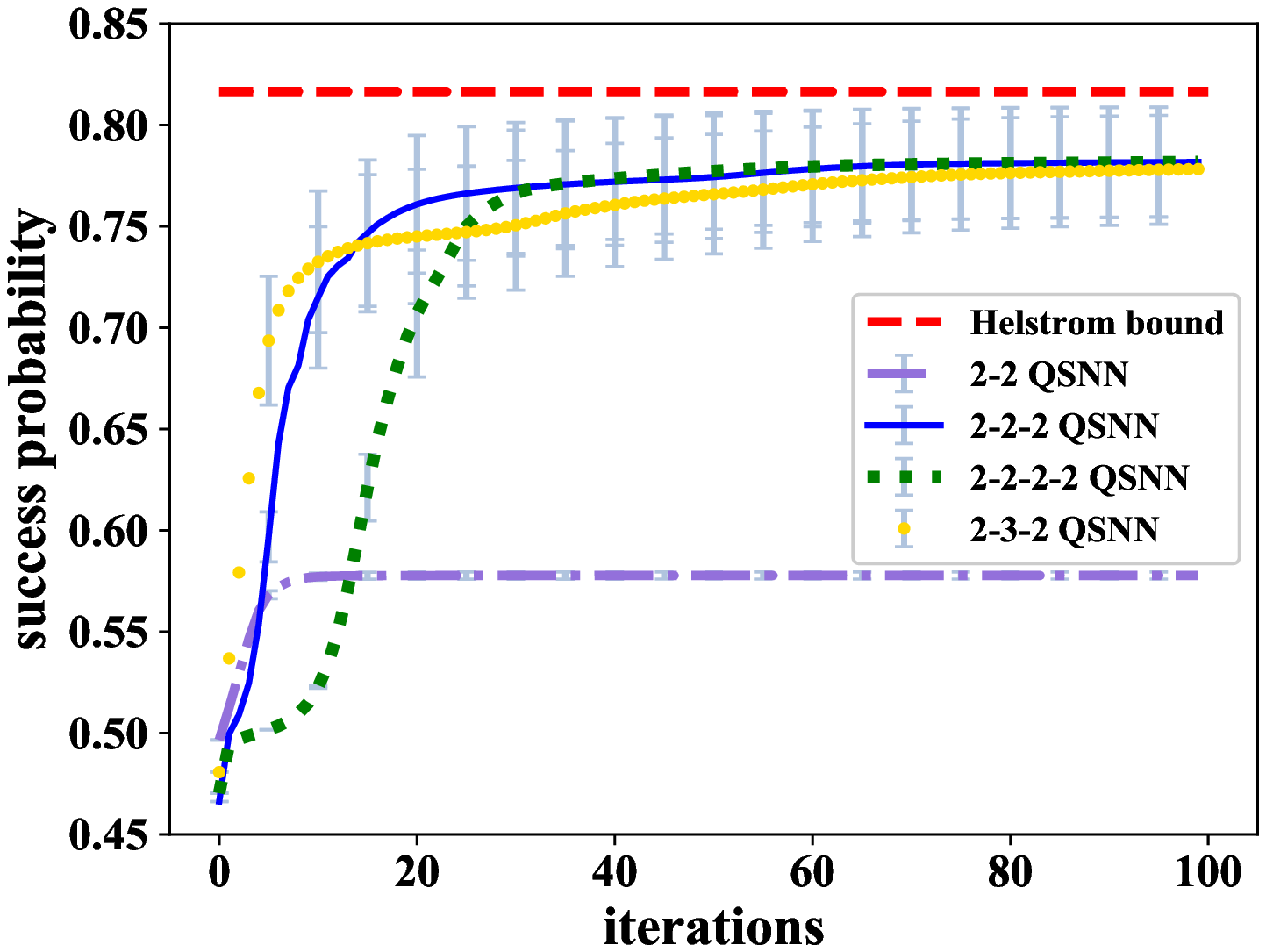}
  \caption{The training performances of the 4 QSNNs (i.e, 2-2, 2-2-2, 2-3-2, and 2-2-2-2 QSNN) in the discrimination of quantum state pairs $(|\psi_0\rangle, |\psi_{\varphi}\rangle)$.
  The results are given by taking the average over the success probabilities for all training sets with different $\varphi$.
  The red dashed line shows the average optimal theoretical value given by the minimum error discrimination.
  The other lines in different colors and shapes represent the average success probabilities given by the 4 QSNNs.
  The 2-2 QSNN performs much worse than the other three.
  There is little difference in the performances of the 2-2-2, 2-3-2, and 2-2-2-2 QSNN.
  Considering that more hidden layers or more hidden layer neurons in a network will increase the consumption of resources, the 2-2-2 network is appropriate for this task.
  }\label{fig:hidden layer}
 \end{figure}
    First, because Fig. 3(c) and (d) of the main text show that there is a gap between the success probability given by the 2-2-2 QSNN and the theoretical optimum in discriminating two pure states represented by Eq. (11) of the main text, we take this task as an example to investigate whether the 2-2-2 QSNN used in the main text is appropriate or not.
    We train the 2-2, 2-3-2, and 2-2-2-2 QSNN with the training sets  $\{(|\psi_0\rangle, state1), (|\psi_{\varphi}\rangle, state2)\}$ which are different in $\varphi=0, \frac{\pi}{6}, \frac{2\pi}{6}, \frac{3\pi}{6}, \cdots, \frac{11\pi}{6}$.
    We give the plot of the success probability with respect to the number of iterations used in the training procedure, as shown in FIG.~\ref{fig:hidden layer}.
    The 2-2 QSNN performs much worse than the other three QSNNs.
    And there is little difference in the performances of the 2-2-2, 2-3-2, and 2-2-2-2 QSNN.
    Considering that more hidden layers or more hidden layer neurons in a network will increase the consumption of resources, the 2-2-2 network is appropriate for this task.

  \begin{figure}[ht]
  \centering
  \includegraphics[width=10.5cm]{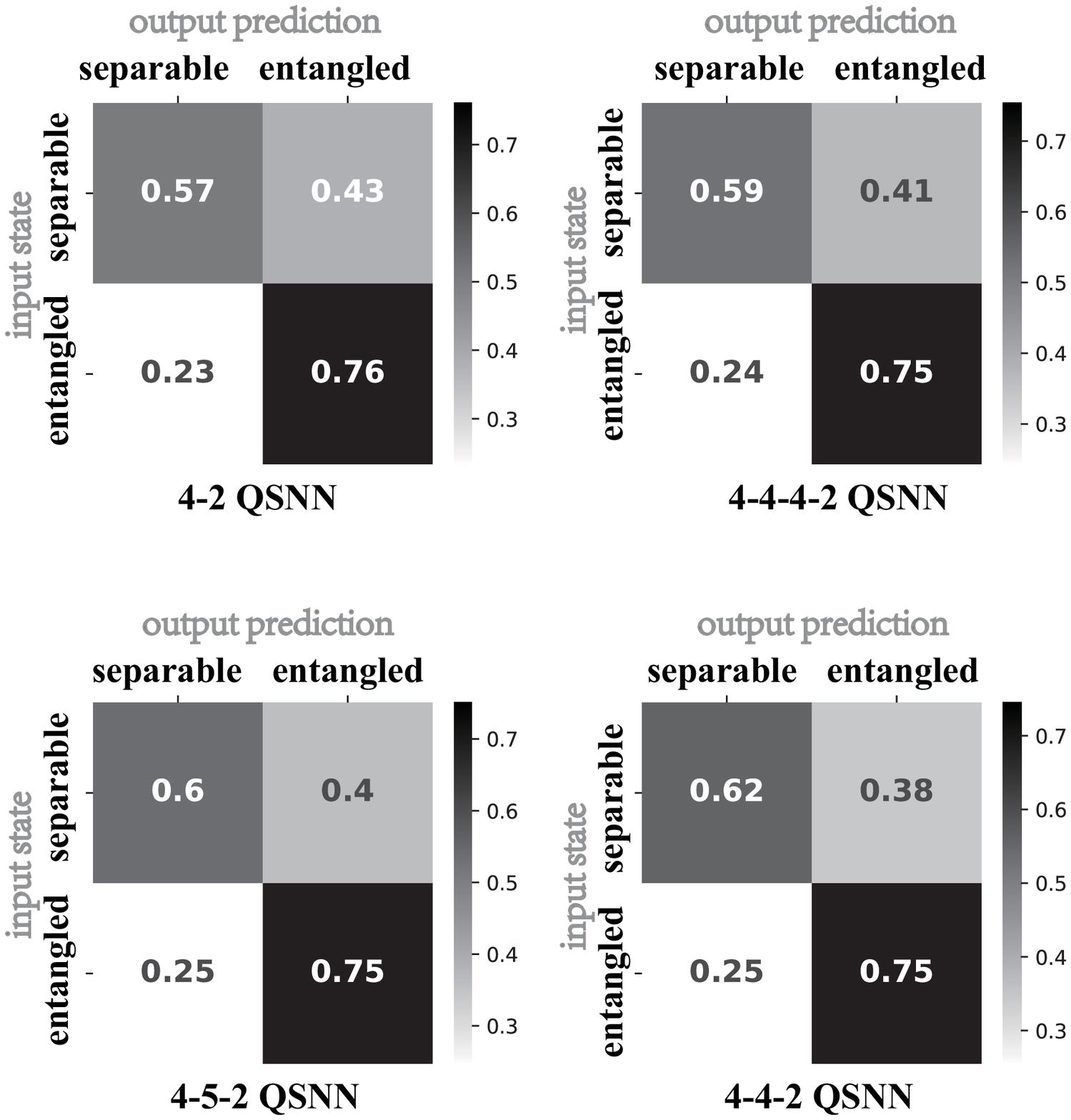}
  \caption{The classification confusion matrices of the 4-2, 4-4-4-2, 4-5-2, and 4-4-2 QSNN.
    The numbers in a classification confusion matrix are the mean success and error probabilities of the QSNN classifying the 49 samples represented by Eq.~(12) (main text) with $p\in\{0.02\cdot n\}_{n=1}^{49}$.
    Compared to the 4-4-2 QSNN used in the main text, the performances of the 4-4-4-2 and 4-5-2 QSNN are not greatly improved, and the 4-2 QSNN performs poorly in identifying separate states.}\label{fig:hidden layer seperate}
 \end{figure}

    Second, we demonstrate that the 4-4-2 QSNN used in the classification task described in the main text \emph{Results –- Classification of Entangled and Separable States} is appropriate.
    We train the 4-2, 4-4-4-2, and 4-5-2 QSNN with the training set as the same as that in \emph{Results –- Classification of Entangled and Separable States}, and use the same test set to evaluate the performance of the QSNNs.
    We show the performances of the 4 QSNNs in FIG.~\ref{fig:hidden layer seperate}.
    Compared to the 4-4-2 QSNN used in the main text, the performances of the 4-4-4-2 and 4-5-2 QSNN are not greatly improved, and the 4-2 QSNN performs poorly in identifying separate states.

    In summary, the QSNN with a single hidden layer is sufficient for the tasks of our interests.
    In addition, it can also be found that more neurons in the hidden layer also can not improve the performance of the QSNN in the tasks described in the main text.

\section{Gradient Descent}\label{subsection:gradient descent}
    In our work, we use gradient descent to optimize the parameters $\theta$ ($h_k$ or $\gamma_k$) in the QSNN to minimize the loss function.
    During the optimization, the parameters $\theta$ are updated iteratively according to
    \begin{equation}
        \theta^\prime=\theta-\eta\frac{\partial \mbox{Loss}}{\partial\theta},
    \end{equation}
    where $\eta$ is the learning rate.
    According to the chain rule, the gradient of the loss function with respect to $\theta$ can be derived as
    \begin{equation}
    \frac{\partial \mbox{Loss}}{\partial\theta} = \sum_s w_s
    \frac{df(\rho_{\text{out}}^s, |l^s\rangle \langle l^s|)}{d\theta} = \sum_s w_s \frac{df}{d\rho_{\text{out}}^s}\frac{d\rho_{\text{out}}^s}{d\theta},
    \end{equation}
    where $f$ is designed according to different tasks,
    and $d\rho_{\text{out}}^s/d\theta$ can be derived from the master equation Eq.~(1) of the main text. To solve it easier, we can rewrite it as a linear equation
    \begin{equation}\label{eq:cj master equation}
    \frac{d|\rho\rangle}{dt}=\mathcal{L}|\rho\rangle,
    \end{equation}
    where $\mathcal{L}$ is the operator corresponding to Liouvillian superoperator, and $|\rho\rangle$ is the vectorized density matrix according to the map
    \begin{equation}
    \rho=\sum_{ij}\rho_{ij}|i\rangle\langle j|\mapsto |\rho\rangle=\sum_{ij}\rho_{ij}|i\rangle\otimes|j\rangle.
    \end{equation}
    With this representation, the operator $L(A)$ corresponding to the left multiplication of a matrix $A$ can be calculated as
    \begin{equation}
    \begin{aligned}
    A\rho= \sum_{ij}\rho_{ij}A|i\rangle\langle j|\mapsto & \sum_{ij}\rho_{ij}(A|i\rangle)\otimes|j\rangle\\
    &= \sum_{ij}\rho_{ij}(A\otimes I)(|i\rangle\otimes|j\rangle)\\
    &=(A\otimes I)|\rho\rangle=L(A)|\rho\rangle,
    \end{aligned}
    \end{equation}
    which indicates that $L(A)=A\otimes I$.
    Similarly, the operator according to the right multiplication is $R(A)=I\otimes A^T$.
    With $L(A)=A\otimes I$ and $R(A)=I\otimes A^T$, the operator $\mathcal{L}$ can be calculated explicitly as
    \begin{equation}\label{eq:Liouvillian superoperator}
        \begin{aligned}
        \mathcal{L}(\vec{h}, \vec{\gamma})=&-i(H\otimes I-I\otimes H^T)\\
        &+\sum_k\left[L_k\otimes L_k^{*}-\frac{1}{2}(L_k^\dagger L_k)\otimes I-\frac{1}{2}I\otimes (L_k^TL_k^*)\right].
        \end{aligned}
    \end{equation}
    Here, $A^*$, $A^T$, and $A^\dagger$ denote, respectively, the complex conjugate, the transpose, and the adjoint of $A$.
    The final state of the QSNN corresponding to the $s$-th sample can be given by solving Eq. (\ref{eq:cj master equation}) as
    \begin{equation}\label{eq:output state}
    |\rho_{\text{out}}^s\rangle = e^{\mathcal{L}(\vec{h},\vec{\gamma})T}|\rho_{\text{in}}^s\rangle.
    \end{equation}
    The gradient can be given as
    \begin{equation}
    \begin{aligned}
        \frac{\partial|\rho_{\text{out}}^s\rangle}{\partial \theta}&=\frac{\partial e^{\mathcal{L}(\vec{h},\vec{\gamma})T}}{\partial \theta}|\rho_{\text{in}}^s\rangle\\
        &=\left\{\int_0^{T}\left[e^{\mathcal{L}(T-t)}\frac{\partial \mathcal{L}(\vec{h}, \vec{\gamma})}{\partial \theta}e^{\mathcal{L}t}\right] dt\right\}|\rho_{\text{in}}^s\rangle.
    \end{aligned}
    \end{equation}
    The partial deviations of the operator $\mathcal{L}$ with respect to parameters can be obtained from Eq. (\ref{eq:Liouvillian superoperator}) easily as
    \begin{equation}
    \begin{aligned}
         &\frac{\partial\mathcal{L}}{\partial h_k}=-i(\mu_k\otimes I-I\otimes \mu_k^T)\\
         &\frac{\partial\mathcal{L}}{\partial \gamma_k}=2\gamma_k\left[\nu_k\otimes \nu_k^*-\frac{1}{2}(\nu_k^\dagger\nu_k)\otimes I-\frac{1}{2}I\otimes(\nu_k^T\nu_k^*)\right],
    \end{aligned}
    \end{equation}
    where $\mu_k=\frac{\partial H}{\partial h_k}$ and $\nu_k=\frac{\partial L_k}{\partial \gamma_k}$.

\section{Mixed State Discrimination}\label{sec:mixed state}

\begin{figure}[ht]
  \centering
  \includegraphics[width=9cm]{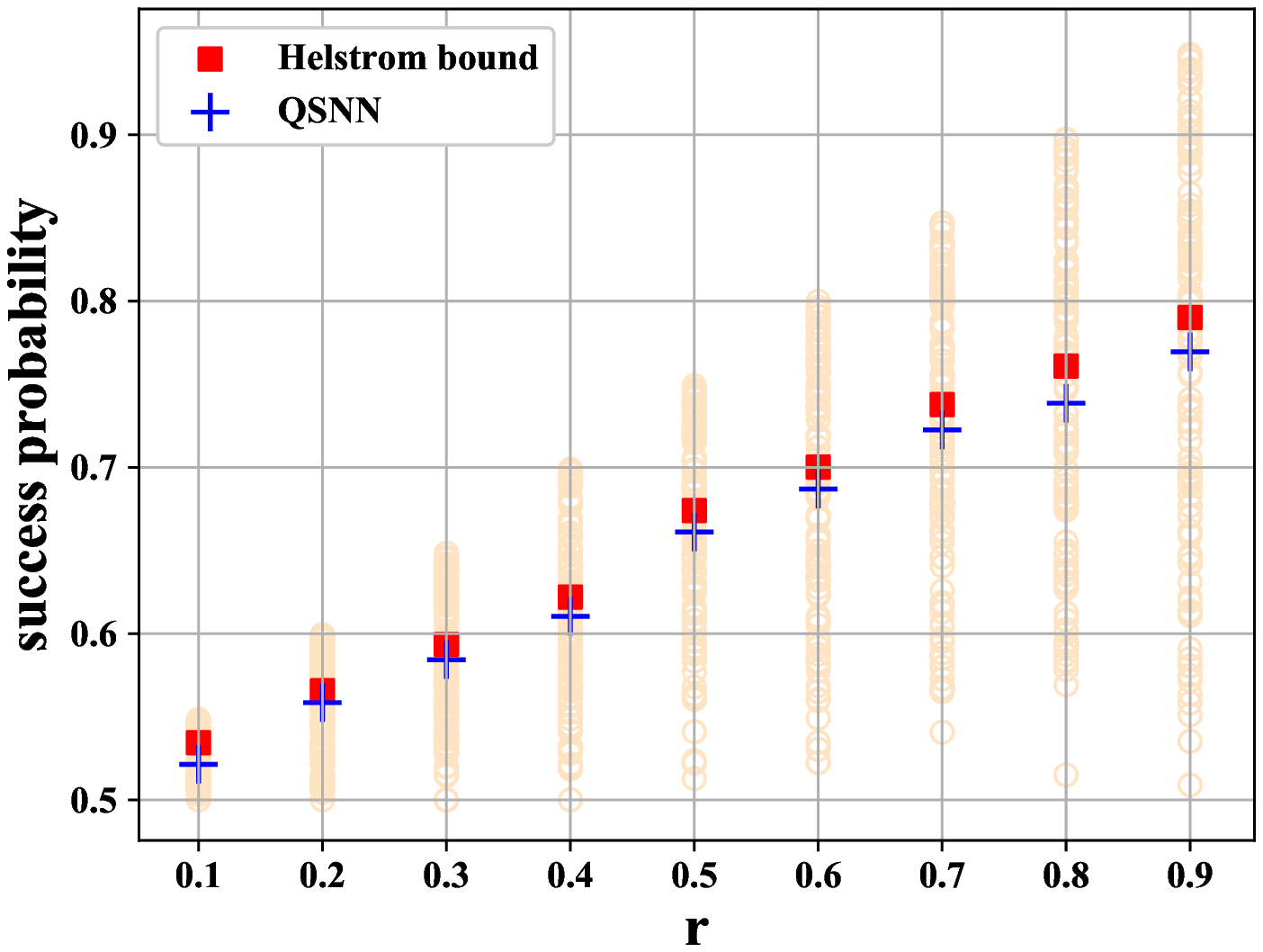}\\
  \caption{The blue crosses represent the average optimal success probabilities of the QSNN in mixed state discrimination, which are compared with the optimal theoretical values (red squares).
  The quantum state pairs are randomly sampled from the sphere surface of radius $r$, and the QSNN's optimal success probability for each pair is shown as an orange circle.
  The performance of the QSNN is close to the ideal value given by the ME discrimination.}\label{fig:mixed}
\end{figure}

    In this part, we investigate the performance of the QSNN in discriminating two mixed quantum states.
    The training set $\{(\rho_{\theta_1,\varphi_1, r}, state1), (\rho_{\theta_2,\varphi_2, r}, state2)\}$ contains two mixed states, and each of them can be given as
    \begin{equation}\label{eq:mixed state}
    \rho_{\theta, \varphi, r} = \frac{I}{2} + \frac{r}{2}(\sin{\theta}\cos{\varphi}\hat{\sigma_x} + \sin{\theta}\sin{\varphi}\hat{\sigma_y} + \cos{\theta}\hat{\sigma_z}).
    \end{equation}
    $I, \hat{\sigma_x}, \hat{\sigma_y}, \hat{\sigma_z}$ are Pauli matrices.
    We randomly choose 100 states pairs $(\rho_{\theta_1,\varphi_1, r}, \rho_{\theta_2,\varphi_2, r})$ on the sphere surface of radius $r$.
    We show the numerical result in FIG.~\ref{fig:mixed}.
    For each value of $r$, the optimal success probability of the QSNN for each training set is shown as an orange circle, and the mean of them is shown as a blue cross.
    The average values of the Helstrom bound are shown as red squares.
    We can see that the performance of the QSNN is close to the Helstrom bound.
    With the decrease of $r$, the quantum states are harder to be discriminated against, which is intuitive because the distance between two states on a smaller sphere is smaller.

\section{Multi-State Discrimination}\label{multiple state}
    In traditional methods, one can construct a measurement to discriminate between two quantum states.
    There are two strategies to design the optimal measurement, i.e., the minimum error discrimination and the unambiguous discrimination.
    However, in general, if the number of states to be discriminated against is greater than two, the optimal measurement is hard to be analytically developed in the fashion of traditional strategies.
    In opposite, our approach is not limited by the number of states theoretically.

    In this section, we take 1 qubit pure state discrimination as an example to show the performance of our QSNN in discriminating multiple quantum states.
    The QSNN we used here consists of 2 neurons in the input layer.
    The 1 qubit states to be processed are encoded in the states of the 2 neurons in the input layer.
    The number of neurons in the output layer is equal to the number of states to be discriminated against.
    The QSNN contains a single hidden layer, where the number of neurons is the same as that in the output layer.
    The connections between neurons are similar to those shown in Fig. 1 of the main text.
    Namely, the Lindblad operators only connect the neurons in the adjacent layers, and the Hamiltonian couplings only exist between some neurons in the input and hidden layer.
\begin{figure}
  \centering
  \includegraphics[width=9cm]{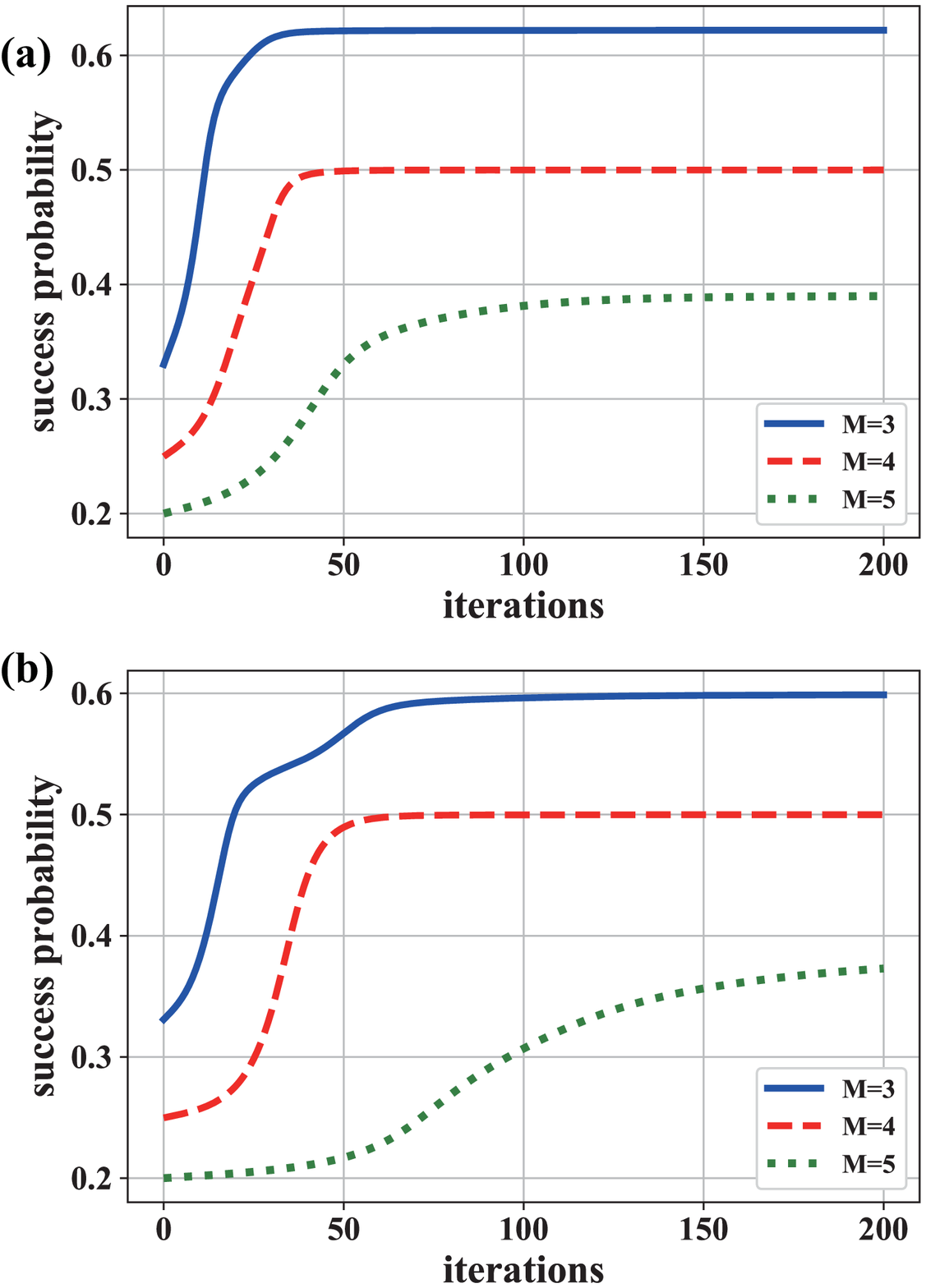}
  \caption{We train the QSNN to discriminate $M$ pure states in the real vector space (see a) or in the complex space (see b). It is a plot of the success probability of the QSNN with respect to the number of iterations used in the training procedure. As the number of states increases, there is a certain decrease in the success probability of the QSNN.
  However, the success probability increases with the number of iterations used in the training procedure, so our approach based on training is somewhat helpful for multiple quantum state discrimination.
    }\label{fig:multi state}
 \end{figure}

    First, we consider discriminating $M$ pure states in the real vector space, and $\{|0\rangle, |1\rangle\}$ is an orthogonal basis of the space.
    The $s$-th state can be given as
    \begin{equation}
        |\psi_s\rangle=\cos{\frac{2\pi s}{M}}|0\rangle+\sin{\frac{2\pi s}{M}}|1\rangle,
    \end{equation}
    which corresponds to a label $l^s\in\{N-s\}_{s=1}^M$.
    After the initialization process, the QSNN evolves according to Eq. (1) of the main text, and gives the final state $\rho_{\text{out}}^s$.
    The prior probabilities $w$ of all the states are equal, i.e., $w = \frac{1}{M}$.
    Performing a projective measurement $\Omega=|l^s\rangle\langle l ^s|$ on the final state $\rho_{\text{out}}^s$ gives
    \begin{equation}
        P_N^s=\text{Tr}(\rho_{\text{out}}^s\Omega^s),
    \end{equation}
    which is the probability that the QSNN gives the label of the $s$-th input correctly.
    So, similar to Eq. (9) of the main text, the loss function can be given as the distance between 1 and the average success probability of the network, that is
    \begin{equation}
        \mbox{Loss} = 1-\frac{1}{M}\sum_{s=1}^M P_N^s.
    \end{equation}
    When the number of states $M$ is 3, 4, or 5, we give the plot of the average success probability with respect to the number of iterations used in the training procedure in FIG.~\ref{fig:multi state}(a).

    Second, we consider discriminating $M$ pure states in a complex space, where the $s$th state can be represented as
    \begin{equation}
        |\psi_s'\rangle=\frac{\sqrt{2}}{2}(|0\rangle+e^{\frac{i2\pi s }{M}}|1\rangle).
    \end{equation}
    We also give the plot of the average success probability of the network with respect to the number of iterations, when the number of states $M$ is 3, 4, or 5, in FIG.~\ref{fig:multi state}(b).

    As the number of states increases, there is a certain decrease in the success probability of the QSNN.
    However, the success probability increases with the number of iterations used in the training procedure, as shown in FIG.~\ref{fig:multi state}(a) or (b).
    So, our approach based on training is somewhat helpful for the multi-state discrimination that is hard to be completed by traditional strategies.

\section{Parameters and Hyperparameters}\label{sec:parameter}
    In this part, we provide some training details about the tasks in the main text, i.e., the initial and final values of the parameters, and the values of hyperparameters.
\begin{figure}
  \centering
  \includegraphics[width=10cm]{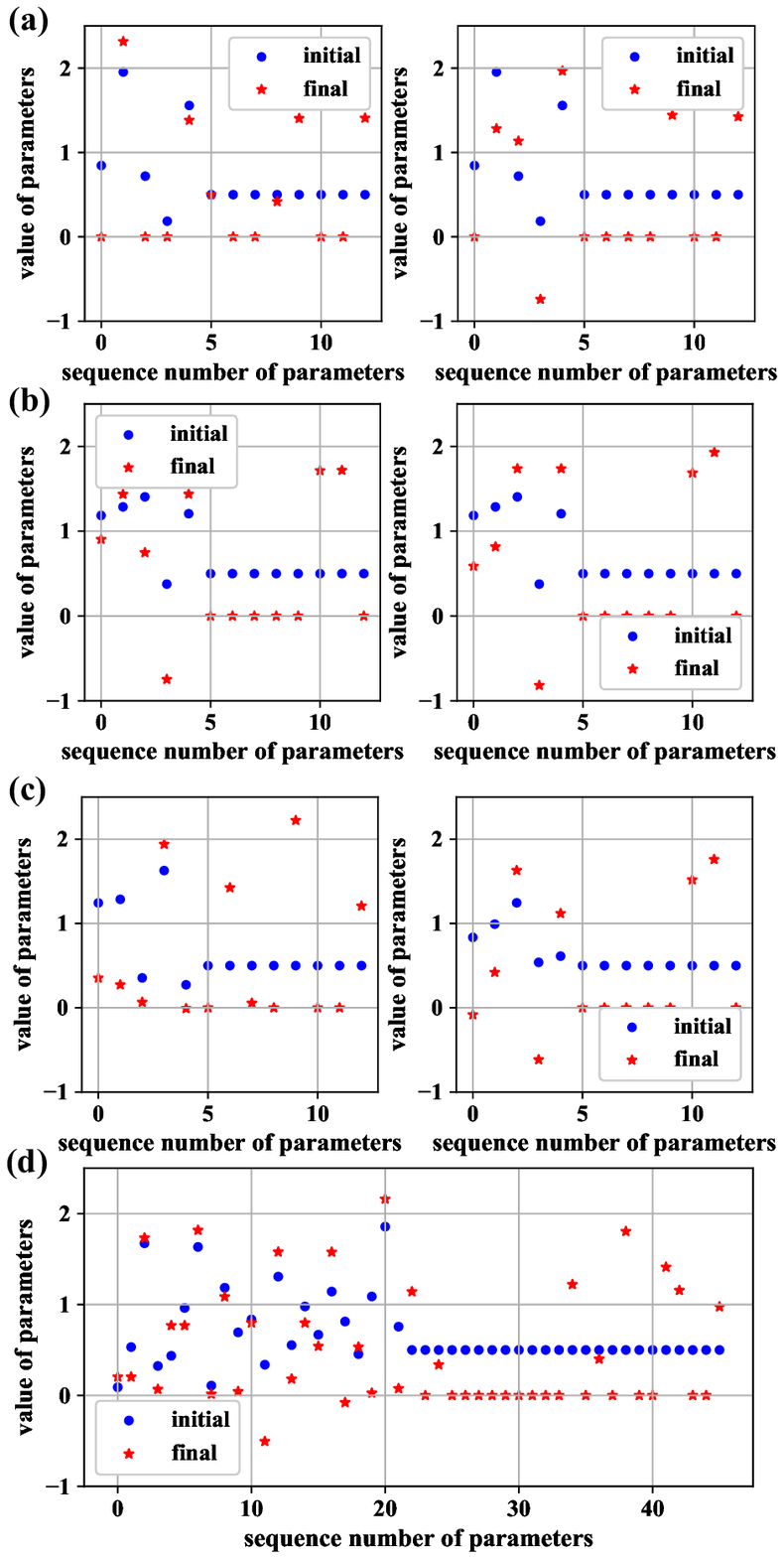}
  \caption{A plot of the initial (blue dots) and final values (red stars) of the network parameters in the 4 tasks, i.e., (a) discriminating two pure states represented by Eq. (10) (main text), (b) discriminating two pure states represented by Eq. (11) (main text), (c) discriminating two mixed states represented by Eq. (\ref{eq:mixed state}), and (d) the classification task described in \emph{Results –- Classification of Entangled and Separable States} of the main text.
  The abscissa of the figure represents the sequence number of the parameters, and the ordinate of the figure represents the value of the parameters.
  In the state binary discrimination task (see a, b, and c), the first 5 parameters are coherent strengths $h_k$ in the Hamiltonian, and the last 8 parameters are the dissipation rates $\gamma_k$ in the Lindblad operators.
  In the classification task (see d), the first 22 parameters are coherent strengths $h_k$ in the Hamiltonian and the last 24 parameters are the dissipation rates $\gamma_k$ in the Lindblad operators.}\label{fig:param}
\end{figure}

    First, we give the initial and final values of the parameters used in each task of the main text.
    In our approach, the parameters that need to be optimized are the coefficients in the Hamiltonian and Lindblad operators, namely $\vec{h}$ and $\vec{\gamma}$.

    \emph{Pure states discrimination}:
    In \emph{Results –- Quantum State Binary Discrimination} of the main text, we discriminate two pure states represented by Eq.~(10) (main text) by training the QSNN with the training set $\{(|\psi_0\rangle, state1), (|\psi_{\theta}\rangle, state2)\}$.
    We separately use 12 training sets with different $\theta$ to train the QSNN.
    The training results are shown in Fig.~3(a) and (b) of the main text.
    There are 12 sets of parameters corresponding to the 12 training sets.
    We choose 2 sets of parameters from them randomly and show their initial and final values in FIG.~\ref{fig:param}(a).
    Then, we discriminate two states represented by Eq.~(11) (main text) by training the QSNN with 12 training sets $\{(|\psi_0\rangle, state1), (|\psi_{\varphi}\rangle, state2)\}$.
    We separately use 12 training sets with different $\theta$ to train the QSNN.
    The training results are shown in Fig.~3(c) and (d) of the main text.
    There are also 12 sets of parameters corresponding to the 12 training sets.
    We choose 2 sets of parameters from them randomly and show their initial and final values in FIG.~\ref{fig:param}(b).

    \emph{Mixed states discrimination}:
    In Sec. \ref{sec:mixed state}, we discriminate two mixed states represented by Eq.~(\ref{eq:mixed state}) by training the QSNN with the training set $\{(\rho_{\theta_1,\varphi_1, r}, state1), (\rho_{\theta_2,\varphi_2, r}, state2)\}$.
    We separately use 100 training sets with different $\theta$, $\varphi$ and the same $r$ to train the QSNNs.
    And the radius $r$ can be chosen from ${0.1, 0.2, \cdots, 0.9}$.
    The result is shown in FIG.~\ref{fig:mixed}.
    There are 900 sets of parameters corresponding to the 900 random training sets.
    We choose 2 sets of parameters from them randomly and show their initial and final values in the FIG.~\ref{fig:param}(c).

    As shown in FIG.~\ref{fig:param}(a), (b), and (c), in the state binary discrimination task, the first 5 parameters are coherent strengths $h_k$ in the Hamiltonian, and the last 8 parameters are the dissipation rates $\gamma_k$ in the Lindblad operators.

    \emph{Classification of entangled and separate states}:
    In our simulation of \emph{Results –- Classification of Entangled and Separable States} in the main text, we train the QSNN to classify an unknown set of Werner-like states represented by Eq.~(12) of the main text, for the case that two local unitaries are fixed as $U_1=\sigma_z,\ U_2=I$.
    The initial and final values of the parameters are shown in FIG.~\ref{fig:param}(d), where the first 22 parameters are coherent strengths $h_k$ in the Hamiltonian, and the last 24 parameters are the dissipation rates $\gamma_k$ in the Lindblad operators.

    Second, we give the values of the hyperparameters used in each task of the main text.
    A hyperparameter is a parameter whose value is determined before training to control the training process.
    One of the hyperparameters is the time $T$ it takes for the network to complete an evolution.
    In our approach, the evolution time $T$ is considered dimensionless (it is of dimension $1/\gamma$ actually, where $\gamma$ is a typical value of the coupling parameters $h_k$ in Hamiltonian or the dissipation rates $\gamma_k$ in Lindblad operators).
    We fix time $T$ to find the optimal values of parameters ($h_k$ and $\gamma_k$) in the network.
    In principle, the value of time can be set to any number.
    However, in practical implementation, the adjustable range of parameters is often determined by the system used to simulate the network.
    One can adjust the time $T$ to make optimal values of all the parameters fall within the range. In all our simulations of the tasks, we empirically fix $T$ as 10 to make the values of parameters in the order of magnitude 1.
    It can be seen from FIG.~\ref{fig:param}, $\gamma T$ is in the order of magnitude 10.
    The other hyperparameter is the learning rate $\eta$.
    It controls the update step sizes of the parameters during training.
    The value of $\eta$ is set as 10 in the training process shown in Fig. 3(a) (main text), and is set as 20 in the other tasks in the main text.

\section{3-Qubit State Discrimination}
\begin{figure}[h]
\centering
\includegraphics[width=9.0cm]{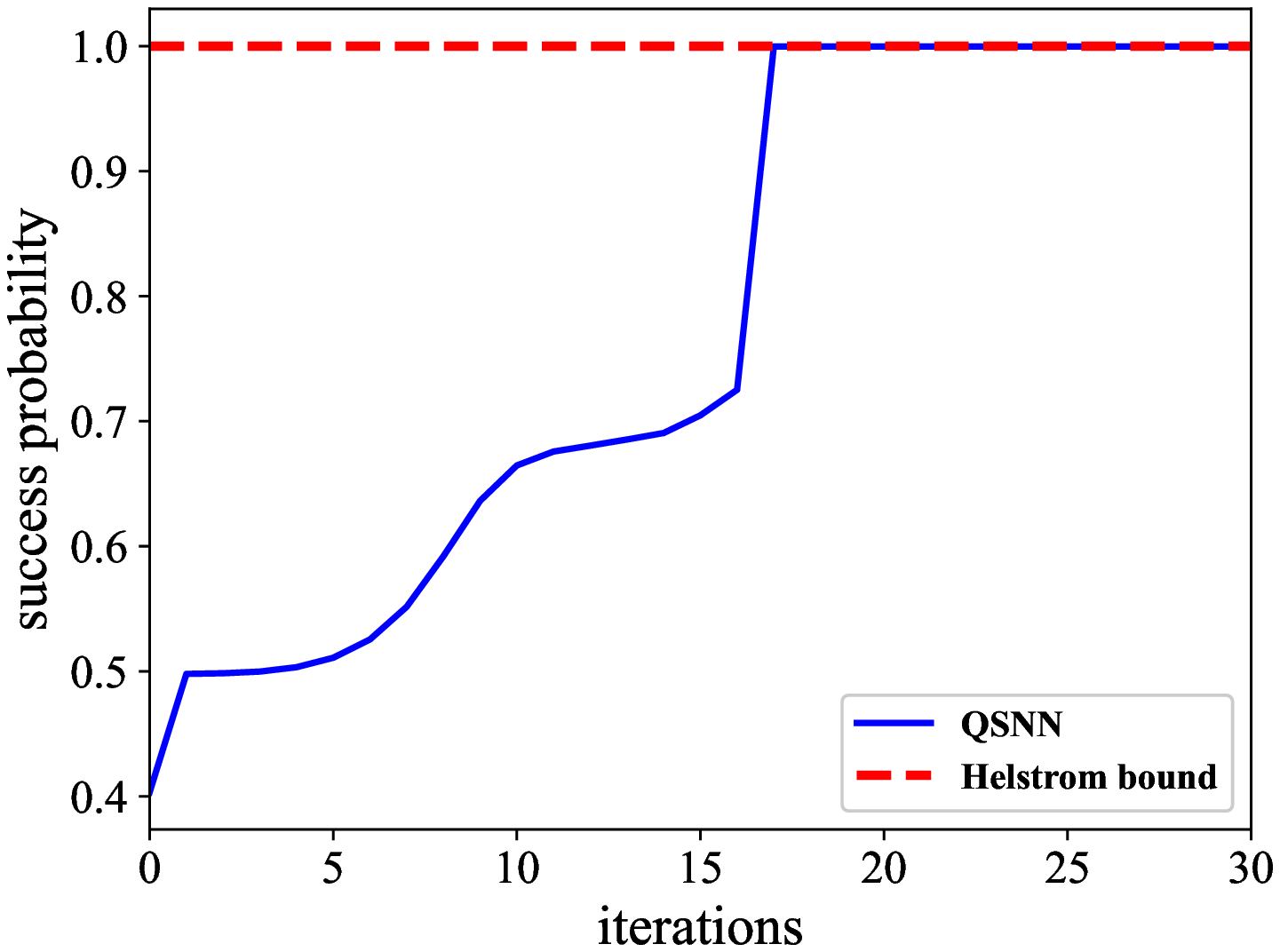}
\caption{The blue curve represents the success probability of the QSNN for the discrimination of the quantum state pairs $(|GHZ\rangle, |W\rangle)$.
  The red dashed line shows the optimal theoretical value given by the minimum error discrimination.
  The success probability of the QSNN rises rapidly, and after less than 20 iterations, it achieves the Helstrom bound.}\label{fig:3qubit}
\end{figure}

Here, we extend the simulation of state binary discrimination (discussed in \emph{Result -- Quantum State Binary Discrimination}) to the unknown 3-qubit states case. The example states here are chosen as two important entanglement states called Greenberger–Horne–Zeilinger (GHZ) state
\begin{equation}
    |GHZ\rangle = \frac{1}{\sqrt{2}}(|000\rangle+|111\rangle)
\end{equation}
and W state
\begin{equation}
    |W\rangle = \frac{1}{\sqrt{3}}(|001\rangle+|010\rangle+|100\rangle)
\end{equation}
for 3 qubits.
In order to encode the 3-qubit input states, the network we used here contains 8 neurons in the input layer, 2 neurons in the hidden layer, and 2 neurons in the output layer.
The training result is shown in FIG.~\ref{fig:3qubit}. The success probability of the QSNN rises rapidly, and after several iterations, it achieves the Helstrom bound.
It shows that our model also works for 3-qubit states.
So, we have confidence that this model is scalable when experimental or simulation conditions allow.

\section{Flow Charts of the Classification of Entangled and Separable States}
\begin{figure}[h]
\centering
\includegraphics[width=9.0cm]{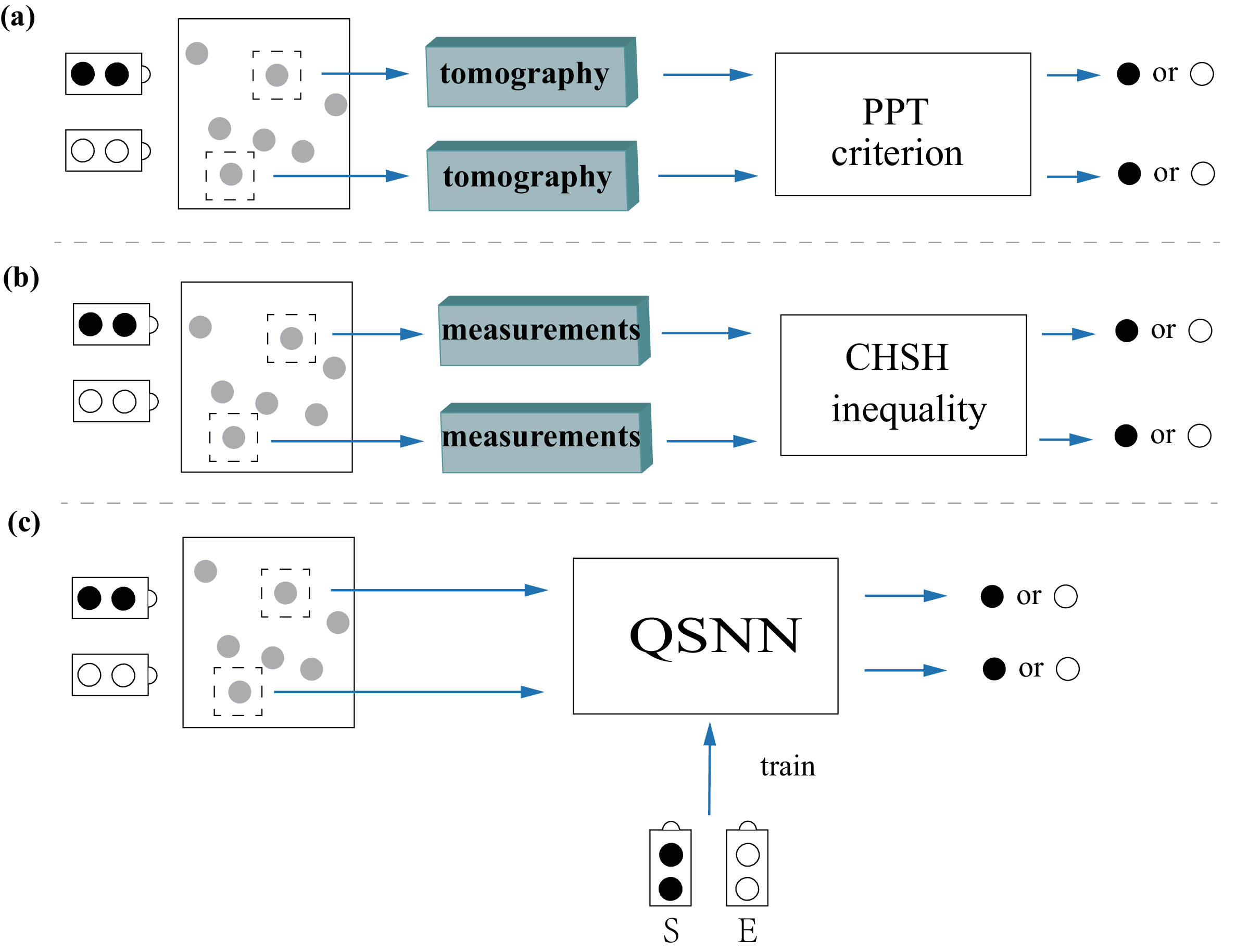}
\caption{The flow charts of three methods of the classification of entangled and separable states.
There are two devices, one of them produces different unknown entangled quantum states (black dots), and the other produces unknown separable states (white dots). The quantum states prepared by them are then mixed in the ensemble (gray dots). The classification task is to determine which device each state is produced by.
(a) Entanglement detection using the PPT criterion. For each quantum state, tomography is needed before the determination.
(b) Entanglement detection using the CHSH inequality. Multiple measurements are inevitable in this approach because of the calculation of the left-hand side of the CHSH inequality.
(c) QSNN classification. The QSNN can be trained by some samples respectively prepared by two devices. Then, the trained network can classify each unknown state with a single detection.
}\label{fig:classificationflow}
\end{figure}

In this part, we show the flow charts mentioned in \emph{Results -- Classification of Entangled and Separable States} as FIG.~\ref{fig:classificationflow}.
